\documentclass[a4paper]{SIGEport}
\usepackage{graphicx}
\usepackage{geometry}                                   
\usepackage{caption}
\usepackage[caption=false, font=footnotesize]{subfig}

\usepackage{hyperref}
\hypersetup{
    colorlinks=true,
    citecolor=red,
    linkcolor=red,
    filecolor=red,      
    urlcolor=red,
    pdfpagemode=FullScreen,
}

\begin{document}

\title{ASA-SimaaS: Advancing Digital Transformation through Simulation Services in the Brazilian Air Force}


\author{Joao~P.~A.~Dantas$ ^1 $, Diego~Geraldo$ ^1 $, Andre~N.~Costa$ ^1 $, Marcos~R.~O.~A.~Maximo$ ^2 $, Takashi~Yoneyama$ ^2 $\\
	{\small $ ^1 $Institute for Advanced Studies, Sao Jose dos Campos/SP - Brazil}\\
	{\small $ ^2 $Aeronautics Institute of Technology, Sao Jose dos Campos/SP - Brazil}\\
\thanks{Joao~P.~A.~Dantas, dantasjpad@fab.mil.br; Diego~Geraldo, diegodg@fab.mil.br; Andre~N.~Costa, negraoanc@fab.mil.br; Marcos~R.~O.~A.~Maximo, mmaximo@ita.br; Takashi~Yoneyama, takashi@ita.br. This work has been supported by Finep (Reference nº 2824/20). Takashi~Yoneyama and Marcos~R.~O.~A.~Maximo are partially funded by CNPq – National Research Council of Brazil through the grants 304134/2-18-0 and 307525/2022-8, respectively.	
 }}

\maketitle

\begin{abstract}

This work explores the use of military simulations in predicting and evaluating the outcomes of potential scenarios. It highlights the evolution of military simulations and the increased capabilities that have arisen due to the advancement of artificial intelligence. Also, it discusses the various applications of military simulations, such as developing tactics and employment doctrines, training decision-makers, evaluating new acquisitions, and developing new technologies. The paper then focuses on the Brazilian Air Force's efforts to create its own simulation tool, the Aerospace Simulation Environment (\textit{Ambiente de Simulação Aeroespacial -- ASA} in Portuguese), and how this cloud-based service called ASA Simulation as a Service (ASA-SimaaS) can provide greater autonomy and economy for the military force. The main contribution of this work is to present the ASA-SimaaS solution as a means of empowering digital transformation in defense scenarios, establishing a partnership network, and improving the military's simulation capabilities and competitiveness. \href{http://www.asa.dcta.mil.br/}{[Website]}\footnote{\href{http://www.asa.dcta.mil.br/}{Website: http://www.asa.dcta.mil.br}} \href{https://youtu.be/hVudLd0hRUc}{[Video]}\footnote{\href{https://youtu.be/hVudLd0hRUc}{Video: https://youtu.be/hVudLd0hRUc}}

\end{abstract}


\section{Introduction}

Predicting the outcome of clashes between opposing military forces has been a recurring demand since the formation of the first armies~\cite{biddle1996victory}. Historically, commanders relied on their experience, intuition, and intelligence gathered by their scouts to make strategic and tactical decisions. However, as warfare became more complex and the range of available weapons and tactics expanded, it became increasingly difficult to predict outcomes accurately~\cite{biddle2004military}.

The use of computers in military simulations dates back to the 1950s when the US Army began to use mainframe computers to analyze large datasets and simulate battlefield scenarios~\cite{krepinevich1994cavalry}. Since then, military simulations have become an essential tool for predicting and evaluating the outcomes of potential conflicts.

In recent years, the capabilities of military simulations have expanded significantly thanks to the increased processing power and evolution of artificial intelligence (AI)~\cite{davis2022artificial}. One of the most effective types of simulation is accelerated time simulation, which compresses the time required to simulate a conflict, allowing decision-makers to evaluate the outcome of a potential conflict more quickly~\cite{bae2016accelerated}.

Armed forces use faster-than-real-time simulations for a range of purposes:

\begin{itemize}
\item \textbf{Select courses of action~\cite{sumile2013collaborative}:} Accelerated time simulations enable military decision-makers to explore different courses of action and evaluate their potential outcomes. By simulating various scenarios, commanders can assess the significance of different strategies and make knowledgeable judgments about the best course of action.
\item \textbf{Develop tactics and employment doctrines~\cite{christensen2022agent}:} Simulations provide a platform for developing and refining tactics and employment doctrines, helping military forces identify their strengths and weaknesses and make necessary adjustments to optimize their effectiveness on the battlefield.
\item \textbf{Perform war games to train decision-makers~\cite{schwartz2020ai}:} War games conducted through accelerated time simulations serve as valuable training exercises for military decision-makers. These simulations allow commanders and staff to practice their decision-making skills in realistic scenarios, improving their ability to analyze complex situations, anticipate outcomes, and make effective strategic choices.
\item \textbf{Evaluate new acquisitions~\cite{yuan2020inverse}:} Accelerated time simulations are instrumental in assessing the capabilities and effectiveness of new assets. By simulating the performance of new weapons systems, vehicles, or equipment, military forces can assess their potential impact on the battlefield and make instructed decisions regarding their acquisition and integration into existing forces.
\item \textbf{Develop new technologies~\cite{mittal2020combining}:} Simulation plays a crucial role in developing new technologies, particularly in autonomous systems. For example, AI algorithms for autonomous vehicles must be extensively tested in simulations before being deployed on real platforms. Simulations provide a safe and controlled environment to evaluate the performance and behavior of these technologies, identify potential issues, and refine them before actual deployment.
\end{itemize}

\begin{figure}[!ht]
	\centering
	\includegraphics[width=0.485\textwidth]{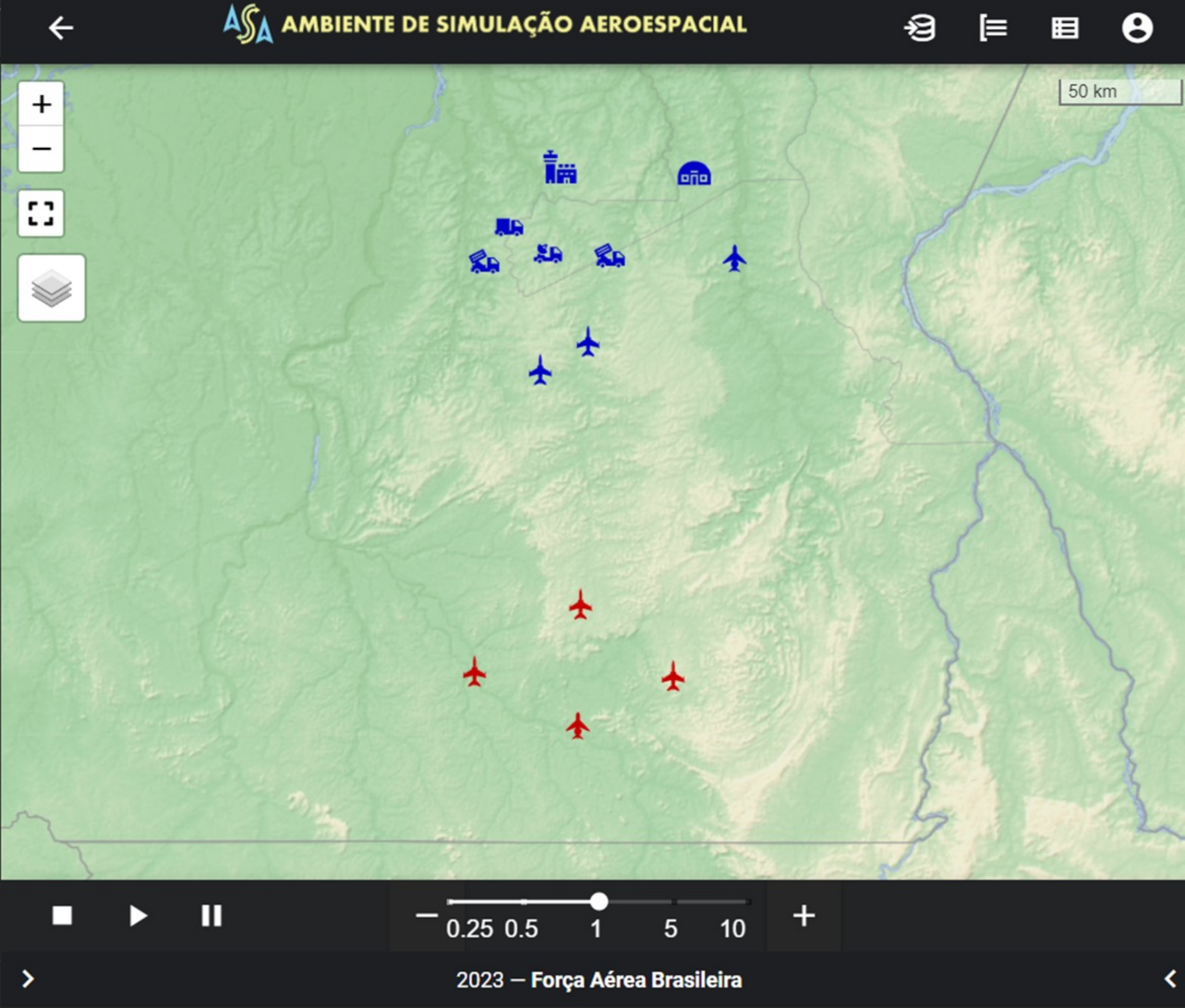}
	\caption{The Aerospace Simulation Environment (ASA) user interface for scenario creation allows users to define multiple simulation aspects, including aircraft models, sensors, communication links, and mission objectives.}
	\label{fig:asa}
\end{figure}

Although many commercial off-the-shelf solutions are available on the market to meet the demand for accelerated time simulations~\cite{lee2020systematic}, in recent years, some armed forces have tried to develop their own simulation tools. The Brazilian Air Force (FAB) is one example of a military force that has taken steps to create its own simulation tool, known as the Aerospace Simulation Environment (\textit{Ambiente de Simulação Aeroespacial -- ASA} in Portuguese), a custom-made simulation framework developed by the Institute for Advanced Studies, that enables the modeling and simulation of military operational scenarios~\cite{dantas2022asa} (Fig.~\ref{fig:asa}). ASA allows the creation, configuration, and execution of defense scenario simulations through a cloud-based service called Simulation as a Service (SimaaS)~\cite{nato}. This new functionality, which is called ASA-SimaaS, could contribute in establishing a partnership network involving government agencies, academic institutions, and companies from the Brazilian Defense Industrial Base (BID), offering means for interoperability, cooperation, and competitiveness in the simulation field within the country. 

Therefore, the main contribution of this paper is to present the ASA-SimaaS solution to empower digital transformation in FAB, by providing a custom cloud-based simulation service for creating, configuring, and executing Defense scenario simulations.

The paper is organized as follows. Section~\ref{sec:process-before} provides a detailed account of the current process of acquiring and maintaining simulation software within the FAB, outlining its challenges and limitations. Section~\ref{sec:effs-effy-indicators} discusses effectiveness and efficiency indicators that can be used to measure the success of the digital transformation process. Section~\ref{sec: asa-management-system} describes a web application that helps manage requests for updates, corrections, and new functionalities in ASA-SimaaS. Section~\ref{sec:advancing-digital-transformation} describes the process after the proposed digital transformation, i.e., the desired end state, focusing on the benefits and potential impact of the ASA-SimaaS on FAB's organizations. Finally, in Section~\ref{sec:conclusions}, we provide some conclusions based on the information presented, highlighting the potential of ASA-SimaaS to positively impact FAB decision-making at all levels and the means of assistance to all elements of the defense triple helix: government, academia, and industry.
\section{The Process Before the Digital Transformation}
\label{sec:process-before}

The initiatives and needs related to simulation within the FAB are managed in a decentralized way, i.e., different organizations have some sort of autonomy to address their simulation demands through specific and individual solutions that are typically acquired from the market.

These solutions often came with a high price tag, requiring significant investment from the owner, both in terms of purchasing the software and providing the necessary training to use it effectively. In addition, these solutions require constant maintenance and updates, which further increase the cost of ownership.

An organization can often face resource unavailability when attempting to acquire simulation software, and those who manage to acquire it can encounter difficulties in maintaining the software lifecycle. For example, license renewals and version updates are often necessary, which requires additional time, effort, and investment.

\begin{figure}[!htp]
	\centering
	\includegraphics[width=0.485\textwidth]{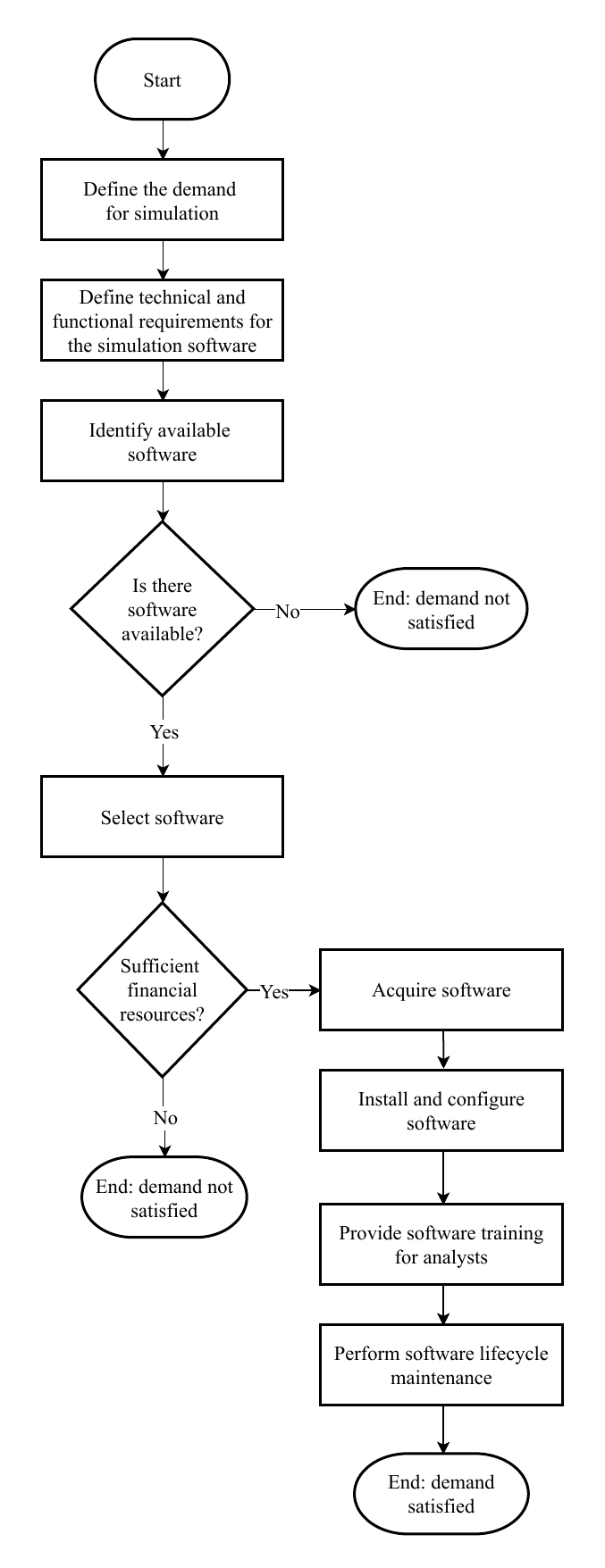}
	\caption{Acquiring Simulation Software: A Typical Process Overview.}
	\label{fig:acquisition}
\end{figure}

The process of acquiring simulation software typically involves several stages, as illustrated in Fig. \ref{fig:acquisition}. First, the organization would identify the demand for simulation software and define the technical and functional requirements. Then, they would search for available options on the market and evaluate them against their requirements. If a suitable solution was found, negotiations for purchase and licensing would occur. Once the software was acquired, the organization would need to install and configure it and then train personnel to use it effectively. Finally, continued maintenance and updates were required to guarantee that the software remained operational and up to date.

Overall, acquiring and maintaining simulation software is a time-consuming, resource-intensive, and expensive process that presents substantial challenges for any organization interested in its use. The ASA-SimaaS can be a reasonable alternative over the previous approach, as it provides a centralized, cloud-based solution that can be accessible to all organizations within the FAB and eliminates many of the burdens associated with traditional software acquisition and maintenance.

\section{Effectiveness and Efficiency Indicators}
\label{sec:effs-effy-indicators}

To facilitate a comprehensive evaluation of the proposed solution in relation to other available alternatives, it is crucial to establish indicators that encompass both the effectiveness and efficiency aspects of the simulation systems.

To correctly understand the proposed indicators, it's important to clarify the term ``Defense Scenario'' or simply ``Scenario''. This term describes a geopolitical and military context in which typical elements of armed conflicts are involved, such as fighter aircraft, tanks, warships, missiles, bombs, radars, artillery, and satellites~\cite{jaiswal2012military}. Some elements, like fighter aircraft, require specialized agents for operation -- for example, a pilot capable of handling an escort mission. In the case of accelerated-time simulations, human figures are modeled through AI tools. It is important to note that the models used in simulations must always undergo verification and validation processes to ensure the results are credible~\cite{hartley1997verification}.

The proposed indicators are meant to evaluate the effectiveness and efficiency of a simulation system that aims to support defense operations. The effectiveness indicator (\texttt{Effs}) measures the rate of meeting the demanded scenarios simulation, while the efficiency indicator (\texttt{Effy}) estimates the cost of simulating a single scenario.

Equation~\ref{eq:1} shows how to calculate the effectiveness indicator based on the number of scenarios simulated (\texttt{NSS}) and the number of scenarios demanded (\texttt{NSD}):

\begin{equation}\label{eq:1}
\texttt{Effs} = \frac{\texttt{NSS}}{\texttt{NSD}}
\end{equation}

The higher the diversity of scenario elements available in the simulation software, the higher the rate of meeting the demanded scenarios simulation. Therefore, simulation systems that allow third-party extensions will perform better in the effectiveness indicator. Aspects such as the availability of resources for software maintenance, as well as trained personnel to operate it, will also impact the effectiveness indicator.

Equation~\ref{eq:2} shows how to measure the efficiency indicator based on the total value invested in the acquisition and maintenance of the system in a given period (\texttt{TVI}) and the number of scenarios simulated using the system in the same period (\texttt{NSS}):

\begin{equation}\label{eq:2}
\texttt{Effy} = \frac{\texttt{TVI}}{\texttt{NSS}}
\end{equation}

The goal of the efficiency indicator is to estimate the cost of simulating one scenario by the system. Ideally, the unit cost value should become lower over time, increasing the positive perception of the return on investment made in the system's acquisition.

It is essential to remark that the effectiveness and efficiency indicators are not standalone metrics and should be analyzed jointly with other parameters, such as the accuracy of the simulation outcomes, the realism of the scenario components, and the usability of the software. Moreover, these indicators should be used as guidelines for decision-making rather than the sole criterion for choosing a simulation system~\cite{junior2021gestao}. The context in which the system will be used, the specific defense objectives, and the available resources should also be taken into consideration when selecting a simulation system.

\section{Asa Management System}
\label{sec: asa-management-system}

\begin{figure}[!htp]
	\centering
	\includegraphics[width=0.485\textwidth]{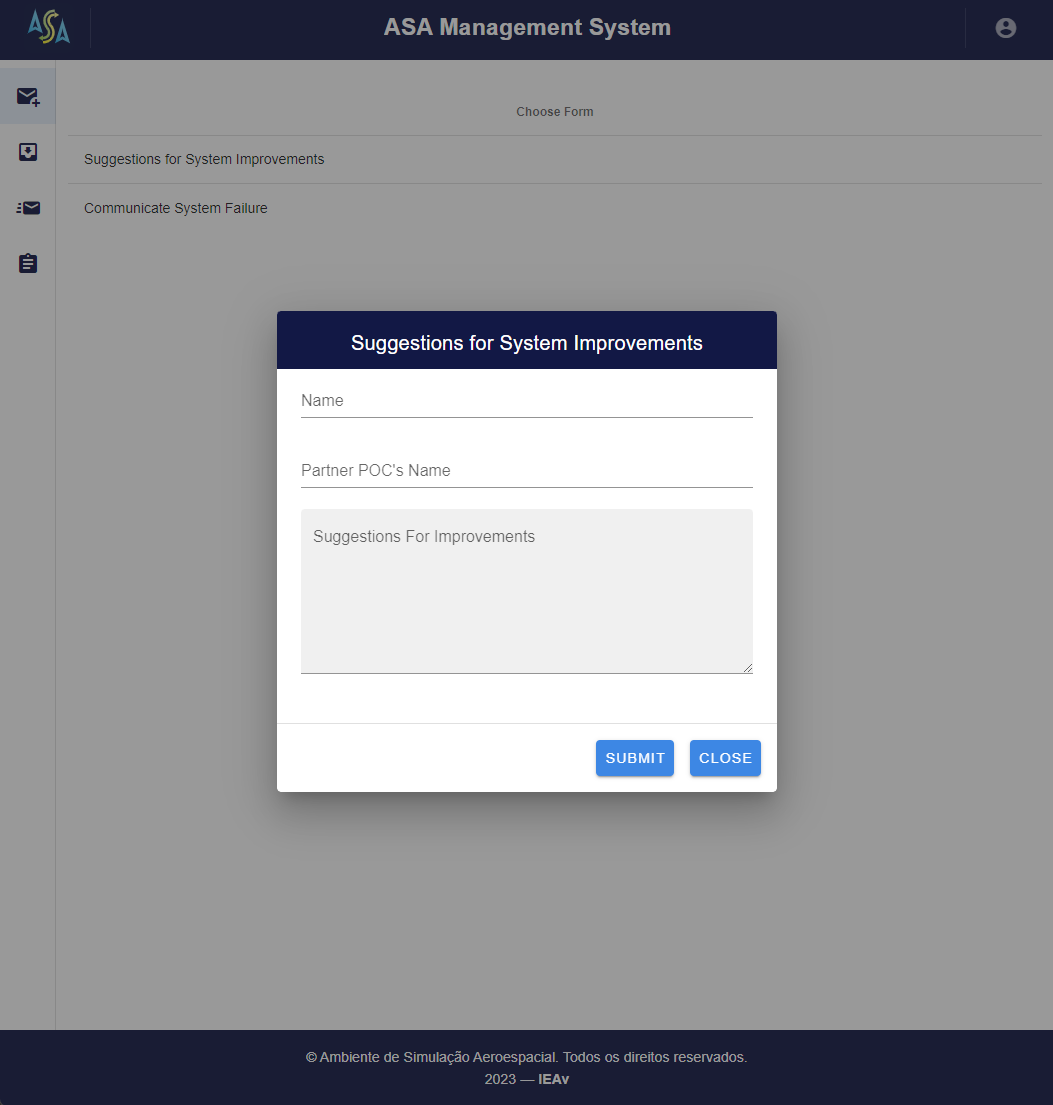}
	\caption{AMS's form used for providing suggestions to generate system improvements.}
	\label{fig:ams}
\end{figure}

The ASA-SimaaS not only offers simulation tools but also incorporates a simple, yet effective, management system that enables standardized communication between users and service managers. This system is designed to ensure that all simulation needs of FAB are adequately cataloged and prioritized, thereby avoiding unnecessary acquisitions and maximizing the number of customers served. Therefore, the Asa Management System (AMS) is a useful web application that facilitates evaluating, prioritizing, and monitoring requests for updates, corrections, and new functionalities within ASA-SimaaS. It empowers service managers to efficiently respond to service access requests and effectively manage the entire lifecycle of the ASA Service. Overall, AMS is designed to coordinate simulation-related initiatives and needs throughout FAB. It promotes a more efficient and collaborative approach to organizational simulation needs management. The example in Fig.~\ref{fig:ams} demonstrates the usage of AMS for system improvement.

The ASA community consists of two main user groups: user-managers and ordinary users. User-managers have the authority to generate and fulfill demands, while ordinary users can only submit demands for consideration. User-managers are further categorized based on thematic areas, including:

\begin{itemize}
\item \textbf{Overall Manager:} coordinating the service, overseeing resource allocation, and acting as the primary Point of Contact (POC) for interactions with military and civil organizations;
\item \textbf{Partnership Manager:} addressing requests from organizations seeking accreditation as partners/users of the service;
\item \textbf{Technical Manager:} evaluating demands that involve changes to the simulation framework or the inclusion/update of models compatible with the ASA;
\item \textbf{Support Manager:} ensuring the correct operation and availability of ASA Service applications, excluding the simulation framework and models;
\item \textbf{Modeling and Simulation Manager:} receiving requests related to scenario simulations, encompassing operational, strategic, tactical needs;
\item \textbf{Intel Manager:} receiving requests related to intelligence information for populating the scenarios; and
\item \textbf{Research and Development Manager:} receiving requests related to research and development areas to support scenario generation and analysis.
\end{itemize}

Within the ordinary users group, several subgroups exist:

\begin{itemize}
\item \textbf{Partner POC:} representing the partner organization and handling institutional demands that benefit all members of the affiliated organization;
\item \textbf{Basic User:} creating scenarios, simulating them, and visualizing their results. This is done through the Web Station, the main ASA user interface;
\item \textbf{Analyst User:} analyzing the data generated by simulations, using the Data Analysis Platform, which facilitates the design of experiments, batch executions, and prediction model generation; and
\item \textbf{Developer User:} adding new features by coding within the service, by accessing the source code repository, which enables the development of the simulation kernel as well as the behavioral, logical, and physical models.
\end{itemize}

\begin{figure}[!htp]
	\centering
	\includegraphics[width=0.485\textwidth]{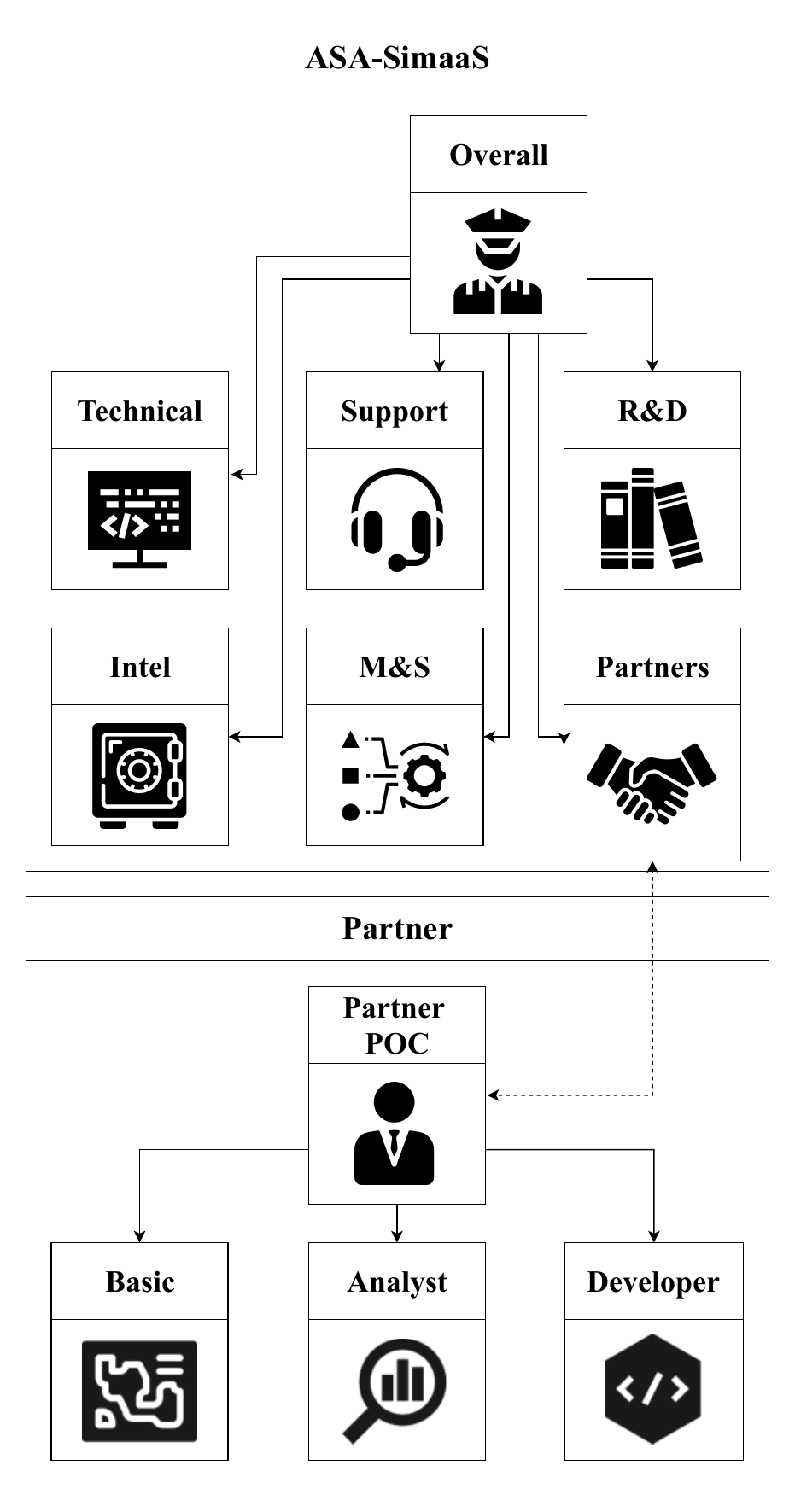}
	\caption{ASA-SimaaS Users: Relationship between distinct user types, each carrying unique permissions that shape their roles.}
	\label{fig:users}
\end{figure}

Each user type (Fig.~\ref{fig:users}) possesses distinct permissions within the simulation service, determining their scope of actions. Additionally, users have customized ``views'' within the AMS interface, allowing non-manager users to create specific demands and managers to evaluate and fulfill them. 

\begin{figure}[!ht]
	\centering
	\includegraphics[width=0.485\textwidth]{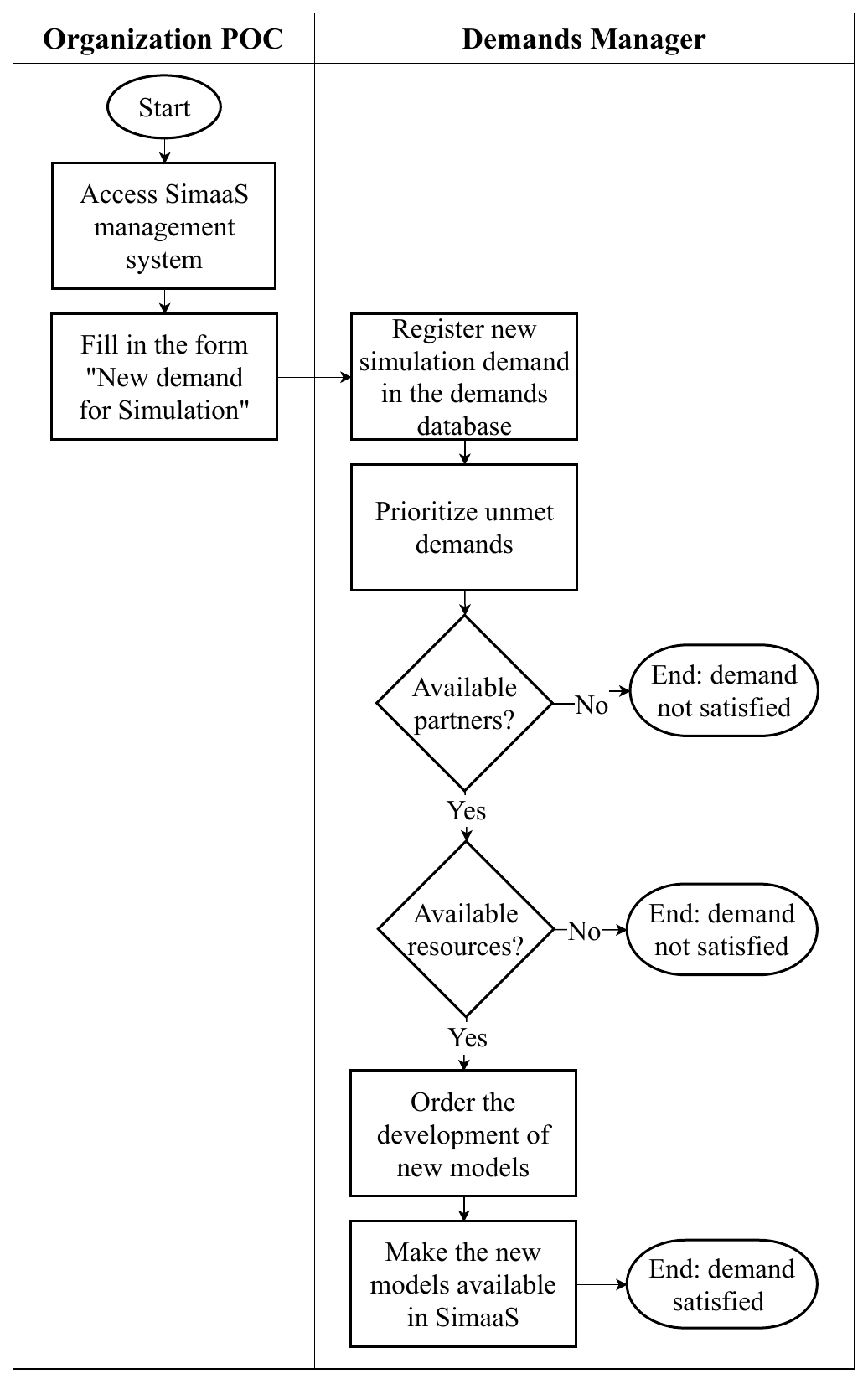}
	\caption{Diagram illustrating the steps involved in generating a report for simulation demand.}
	\label{fig:report}
\end{figure}

The Partner POC represents a special user role since they are responsible for reporting simulation needs that are not yet met by the ASA-SimaaS. For instance, if a certain FAB organization that makes important operational decisions needs to evaluate a scenario that includes an aircraft model that is not yet available in the list of models provided to users, the POC can report their demand through the service management system. The service manager responsible for this type of request will then seek a solution by turning to the network of partners. The flowchart in Fig.~\ref{fig:report} details the process for reporting a simulation demand.

\begin{figure}[!ht]
	\centering
	\includegraphics[width=0.485\textwidth]{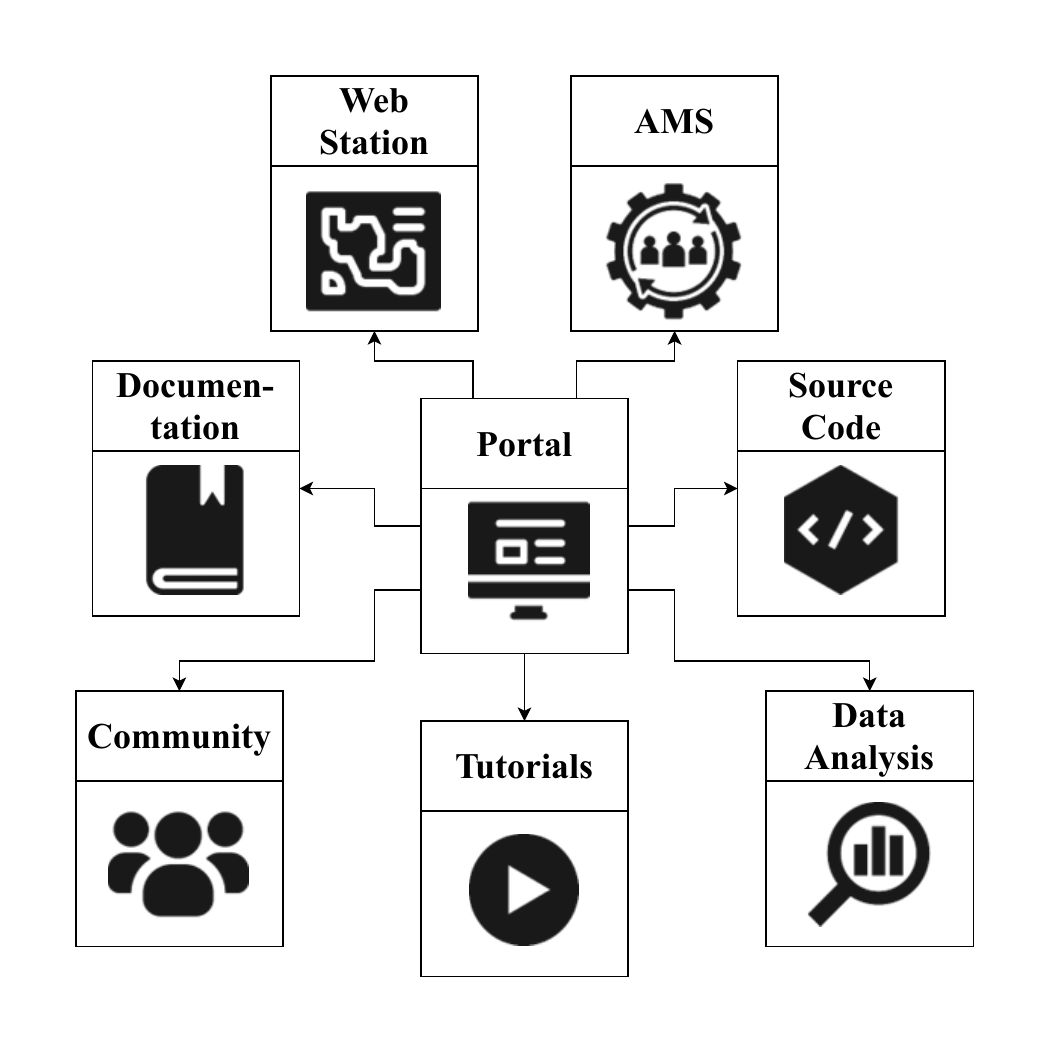}
	\caption{ASA-SimaaS services: aimed at providing comprehensive solutions to assist the simulation service.}
	\label{fig:services}
\end{figure}

The tools and services are further explained and exemplified in the service Documentation and Tutorials modules, which along with the previously mentioned ones are depicted in Fig.~\ref{fig:services}.

\section{Advancing Digital Transformation}
\label{sec:advancing-digital-transformation}

The ASA-SimaaS is expected to have a positive impact on several key areas within FAB, enhancing their capabilities and supporting critical analyses and activities. The following areas are anticipated to benefit from the utilization of this service:

\begin{itemize}
\item \textbf{Chief of Staff:} ASA-SimaaS could provide valuable support for analyses related to Simulation-Based Acquisition (SBA). With the ability to simulate a wide range of scenarios and evaluate the performance of various systems and technologies, a Chief of Staff can make informed decisions regarding future capabilities and acquisitions.

\item \textbf{Operational Commands:} ASA-SimaaS could play a crucial role in supporting the development of doctrines and tactics. By providing a platform for conducting detailed simulations, an Operational Command could analyze different tactical approaches, evaluate their effectiveness, and refine operational doctrines accordingly. ASA-SimaaS also could serve as a valuable tool in the selection of Courses of Action (COA). By simulating various COAs and evaluating their outcomes, one could identify the most effective strategies for specific operational scenarios.

\item \textbf{Command and Staff College:}  War Games are essential for training and evaluating military personnel in realistic scenarios. ASA-SimaaS could be used to support War Game activities and improve the efficacy of training exercises and facilitate the evolution of military personnel's strategic thinking and decision-making skills.

\item \textbf{Research and Development:} ASA-SimaaS could play a special role in supporting the development of Artificial Intelligence (AI) algorithms for a wide range of purposes. By utilizing the simulation environment provided by ASA-SimaaS, researchers can assess and optimize AI algorithms for combat systems and air traffic control applications.

\end{itemize}

The utilization of ASA-SimaaS across multiple organizations can promote collaboration, efficiency, and effectiveness, ultimately contributing to the overall readiness and success of the Brazilian Air Force.

The ASA-SimaaS relies heavily on cloud computing technology to be offered as a web service. Typically, complex simulations such as Defense Scenarios require large computational power for processing. In this case, the simulations demanded via the web are not processed locally on users' computers but in a corporate data center. The data center has a complex Information Technology (IT) infrastructure (hardware and software), available 24/7, that performs demand scaling and parallel execution of simulations. In summary, ASA-SimaaS can enable the organization to perform simulations of complex scenarios at any time of the day without relying on high-powered personal computers
\section{Conclusions}
\label{sec:conclusions}





In conclusion, the introduction of ASA-SimaaS can streamline the digital transformation process of FAB, providing users with a powerful tool for simulating complex defense scenarios and enabling them to optimize the use of high-performance computing resources available in a corporate data center while simultaneously reducing costs associated with software licensing and maintenance.

One of the main features of ASA-SimaaS is its management system, which facilitates seamless communication between users and service managers. This ensures that users can systematically report simulation needs that are not currently addressed. Such information is vital as it helps inform acquisition strategies, enables the prioritization of resources, and ensures that the service evolves in concert with user needs. ASA-SimaaS has the potential to become an integral part of decision-making processes at all levels within FAB, whether they be tactical, operational, or strategic. The robust infrastructure supporting ASA-SimaaS has been meticulously crafted to provide to the defense ecosystem's diverse elements, including government entities, academic institutions, and the defense industry.

Looking toward the future, ASA-SimaaS holds significant potential for enhancement. Improvements in integrating AI could add a new layer of intelligence to the simulations, making them more capable at predicting and adapting to real-world complexities. Another potential development could include real-time collaboration tools, fostering a cohesive environment where users can share insights and make collective, data-driven decisions. Furthermore, integrating ASA-SimaaS with augmented and virtual reality technologies could result in immersive simulation experiences that closely resemble the intricacies of real-life defense scenarios, which could be made available throughout FAB. A critical factor for sustained success will be ensuring that ASA-SimaaS remains scalable and agile, capable of adapting to the ever-changing landscape of defense needs.

In summary, ASA-SimaaS represents a substantial stride in digitizing defense activities, as it empowers FAB to optimize resources, streamline decision-making processes, and enhance support for critical missions. There are high expectations that ASA-SimaaS will positively impact FAB organizations in the years to come, contributing to Brazil's maintenance of a robust and efficient defense posture.

\bibliography{RefSIGE}
\bibliographystyle{IEEEtran}

\end{document}